\documentclass[12pt,english]{article}

\usepackage{lineno}
\usepackage[margin=1in]{geometry}
\usepackage{amstext}
\usepackage{amsfonts}
\usepackage{amsmath}
\usepackage{amsthm}
\usepackage{authblk}
\usepackage{bm}
\usepackage{caption}
\usepackage{color}
\usepackage{epsfig}
\usepackage{float}
\usepackage{gensymb}
\usepackage{graphics}
\usepackage{graphicx}
\usepackage{longtable}
\usepackage{lscape}
\usepackage{mathtools}
\usepackage{mathrsfs}
\usepackage{natbib}
\usepackage{setspace}
\usepackage{times}
\usepackage{titlesec}
\usepackage{verbatim}
\usepackage{fancyhdr}

\titlespacing*{\section}{0pt}{0\baselineskip}{0\baselineskip}
\titlespacing*{\subsection}{0pt}{0\baselineskip}{0\baselineskip}
\titlespacing*{\subsubsection}{0pt}{0\baselineskip}{0\baselineskip}
\floatstyle{plaintop}
\restylefloat{table}
\fancypagestyle{plain}{
  \fancyhf{}
  \fancyhead[C]{Perry J. Williams et al. $\cdot$ Optimal dynamic
    survey designs}
  \fancyfoot[C]{\thepage}
}
\baselineskip=24pt \title{Monitoring dynamic spatio-temporal
  ecological processes optimally}
\author[1,2]{Perry J. Williams
\thanks{Corresponding author: perry.williams@colostate.edu}}
\author[3,2]{Mevin B. Hooten}
\author[4,5]{Jamie N. Womble}
\author[6]{George G. Esslinger}
\author[4]{Michael R. Bower,}

\affil[1]{\small Colorado Cooperative Fish and Wildlife Research Unit,
  Department of Fish, Wildlife, and Conservation Biology, Colorado
  State University, Fort Collins, CO 80523} \affil[2]{\small Department of
  Statistics, Colorado State University, Fort Collins, CO 80523}
\affil[3]{\small U.S. Geological Survey, Colorado Cooperative Fish and
  Wildlife Research Unit,  Department of Fish, Wildlife, and
  Conservation Biology, Colorado State University, Fort Collins, CO}
\affil [4]{\small National Park Service, Southeast Alaska Inventory
  and Monitoring Network, Juneau, AK} \affil [5]{\small National Park
  Service, Glacier Bay Field Station, Juneau, AK} \affil [6]{\small
  U.S. Geological Survey, Alaska Science Center, Anchorage, AK}

\date{}

 \bibpunct{(}{)}{,}{a}{}{;}

\begin{document}
\begin{spacing}{1.8}
  \maketitle
  \pagebreak

  \begin{flushleft}
    \begin{abstract}
      Population dynamics varies in space and time. Survey designs
      that ignore these dynamics may be inefficient and fail to
      capture essential spatio-temporal variability of a
      process. Alternatively, dynamic survey designs explicitly
      incorporate knowledge of ecological processes, the associated
      uncertainty in those processes, and can be optimized with
      respect to monitoring objectives. We describe a cohesive
      framework for monitoring a spreading population that explicitly
      links animal movement models with survey design and monitoring
      objectives. We apply the framework to develop an optimal survey
      design for sea otters in Glacier Bay. Sea otters were first
      detected in Glacier Bay in 1988 and have since increased in both
      abundance and distribution; abundance estimates increased from 5
      otters to $>$5,000 otters, and they have spread faster than 2.7
      km per year.  By explicitly linking animal movement models and
      survey design, we were able to reduce uncertainty associated
      with predicted occupancy, abundance, and distribution. The
      framework we describe is general, and we outline steps to
      applying it to novel systems and taxa.
    \end{abstract}


    \section*{Introduction}

    \setlength{\parindent}{5ex} Population spread is a fundamental
    theme in ecology \citep{bullock2002dispersal}. Applications
    include reintroductions of endangered species, invasive species
    management, and the emergence or re-emergence of wildlife or plant
    disease \citep{hooten2007hierarchical, williams2017integrated,
      hefley2017when}. The distribution and abundance of a spreading
    population is a dynamic process that changes in space and
    time. These dynamics make it challenging to develop efficient
    monitoring designs that must consider, not only where populations
    have been in the past, but also, where populations are expected to
    be in the future. For example, sea otters (\emph{Enhydra lutris})
    in Glacier Bay have increased rapidly in distribution and abundance
    through time, making the development of an efficient monitoring
    design challenging.

    During the multi-national commercial maritime fur trade of the
    18$^{\text{th}}$ and 19$^{\text{th}}$ centuries, sea otters were
    extirpated from southeastern Alaska. Legislation following the
    maritime fur trade, including the International Fur Seal Treaty
    (1911), the Marine Mammal Protection Act (1972), and the
    Endangered Species Act (1977) provided legal protection to sea
    otters from most harvest \citep[Williams et al. \emph{In
      Review}]{kenyon1969sea, bodkin2015historic}. Legal protection,
    combined with translocations by wildlife agencies helped sea
    otters colonize much of their former distribution. By 1988, sea
    otters were documented at the mouth of Glacier Bay. Since then,
    sea otter abundance has increased an estimated 21.5\% per year, a
    rate near their biological maximum reproduction rate. Further, sea
    otters have spread across Glacier Bay at a rate of at least 2.7 km
    per year. They are now one of the most abundant marine mammals in
    Glacier Bay (Williams et al. \emph{In Review}).

    Beginning in 1999, a design-based (probabilistic) survey method
    was used to monitor the abundance of sea otters in Glacier Bay
    \citep{bodkin1999aerial}. The survey was conducted eight times
    between 1999 and 2012, and consisted of systematically selected
    transects with random starting points
    \citep{esslinger2015monitoring}. Survey effort was stratified
    based on ocean depth and shoreline features
    \citep{bodkin1999aerial}. The northern extent of surveys was
    based on the existing distribution of sea otters. Initially, while
    sea otter distribution was relatively concentrated, abundance
    estimates were precise; between 1999 and 2006, the mean of the
    standard errors equaled 280 otters (mean abundance = 1,496).  As
    sea otters increased in abundance and distribution, distance
    between transects were increased to accommodate the increasing
    spatial extent of the sea otter distribution. However, the number
    of transects remained relatively constant due to logistical and
    budgetary constraints. As transects became more sparse, and as
    abundance increased, standard errors of abundance estimates
    increased, as did coefficients of variation. By 2012, the last
    year the survey was conducted, the estimated abundance was 8,508
    sea otters, and the standard error was greater than 2,200 sea
    otters \citep[][Williams et al. \emph{In
      Review}]{esslinger2015monitoring}.

    In 2015, sea otters were selected as a vital sign for long-term
    ecological monitoring by the National Park Service due to their
    role as a keystone apex predator, and their influence in
    structuring nearshore marine communities \citep{estes1974sea}. The
    National Park Service is concerned with developing a statistical
    monitoring framework that maximizes efficiency to estimate sea
    otter abundance and distribution in Glacier Bay. The monitoring
    framework will serve as the foundation for understanding sea
    otters' role as drivers of the nearshore benthic food web. Thus, a
    survey design that provides precise, rigorous, and honest
    estimates of abundance, distribution, and colonization dynamics is
    required.

    Many ecological processes, including population spread, exhibit
    spatial patterns that change over time in a coherent, dynamic
    fashion. These dynamics are often ignored when developing spatial
    survey designs \citep{wikle2005dynamic}. However, efficient
    monitoring of such spatio-temporal systems can be achieved by
    modeling the dynamic system and associated uncertainty, and
    reducing the uncertainty associated with the effect of sampling
    locations at future observation times
    \citep{hooten2009optimal}. There has been a proliferation of
    statistical methods for modeling and forecasting the distribution
    and abundance of a spreading population
    \citep[e.g.,][]{wikle2003hierarchical, wikle2006hierarchical,
      hooten2007hierarchical, hooten2008hierarchical,
      williams2017integrated}. Although mathematical and
    statistical models are ubiquitous for inferring population spread,
    rarely are data collection and modeling explicitly linked in a
    unifying framework.

    Dynamic survey designs provide a cohesive framework for coupling
    models of population spread, and the optimal selection of sampling
    locations. Dynamic survey designs are common in environmental
    monitoring, including: monitoring hurricanes via aircraft
    \citep{wikle1999space}, ozone monitoring \citep{wikle1999space},
    meteorological forecasting \citep{berliner1999statistical}, and
    ground-water-pollution source identification
    \citep{mahar1997optimal}. However, dynamic survey designs have
    been applied to few long-term ecological monitoring programs
    \citep[e.g., ][]{wikle2005dynamic, hooten2009optimal,
      evangelou2012optimal, hooten2012optimal}.

    We have four objectives in this paper: 1) introduce concepts and
    terminology related to optimal dynamic survey designs, 2) describe
    a general statistical framework for mechanistically modeling
    population spread, 3) fuse statistical models of population spread
    and dynamic survey designs in one coherent framework, and 4) apply
    the framework to monitoring sea otters in Glacier Bay. Although we
    motivate this application using monitoring of sea otters in
    Glacier Bay, we describe the methods in sufficient generality to
    be applicable to any system or taxa in which investigators are
    interested in modeling and monitoring the distribution, abundance,
    and colonization dynamics of a spreading population.

    \section*{Optimal dynamic survey design}

    In this section, we describe the general methodology to develop an
    optimal dynamic survey design for a spreading
    population. Population spread is an ecological process that
    evolves spatially through time. To improve our understanding in
    how this process evolves, we first require a baseline
    understanding of the ecological process, and the associated
    uncertainty. Thus, a statistical model that incorporates our
    current understanding of the ecological process is required so
    that we can predict what the behavior of the population is likely
    to do in future monitoring periods. If we can predict future
    behavior, and the associated uncertainty, we can then choose
    survey locations that help reduce uncertainty in our understanding
    of the process \citep{hooten2009optimal}. This is the fundamental
    notion behind the basic steps of dynamic survey designs that we
    describe next.

    Dynamic survey designs can be broken down into a series of steps
    that are each conceptually straightforward
    (Fig. \ref{fig:schematic}). First, a dynamic spatio-temporal
    process, such as occupancy or abundance (and the associated
    uncertainty) is modeled using baseline data.  Second, using the
    model from the first step, a statistical forecast is made. The
    forecast provides a base for examining potential survey designs
    that could be implemented in the future. Third, investigators
    identify the objectives they wish to achieve with their monitoring
    \citep[e.g.,][]{nichols2006monitoring}.  Objectives, or
    \emph{design criteria}, typically include minimizing average
    prediction variance, minimizing maximum prediction variance
    (mini-max), or, the minimizing variance of regression parameter
    estimates \citep{wikle1999space, wikle2005dynamic,
      hooten2009optimal}, but could also include minimizing
    multi-model uncertainty \citep{nichols2006monitoring}, cost
    \citep{hauser2009streamlining}, or some combination thereof
    \citep{williams2017guide}. Fourth, after a design criterion is
    selected, a design is chosen such that it optimizes the design
    criterion. Fifth, data are then collected using the optimal
    design. The original model used to make the forecast is then
    updated with the new data. This process is iterated through time,
    increasing the understanding of the underlying ecological process
    of interest. In this regard, optimal dynamic survey designs are
    analogous to adaptive resource management, an iterative process of
    decision making in the face of uncertainty, with an aim to
    reducing management uncertainty through time by monitoring the
    system's response to management
    \citep[e.g.,][]{johnson1997uncertainty}.

    In what follows, we discuss methods for implementation of these
    steps generally, such that they may be tailored to other systems
    and taxa for which investigators seek to model and monitor
    population spread. We then describe how we tailored these general
    methods to the specific task of modeling and monitoring sea otters
    in Glacier Bay.

    \subsection*{A general spatio-temporal model for population-level
      animal movement}
    Population spread exhibits linear or non-linear dynamics that can
    be classified as \emph{diffusion}. Diffusion refers to the process
    of spreading out over an increasingly larger area through time
    \citep{skellam1951random, wikle2010general}. Partial differential equations (PDE) are
    powerful tools for modeling population-level (i.e., Eulerian)
    animal movement in ecology \citep[e.g., ][]{skellam1951random,
      okubo1980diffusion, andow1990spread, holmes1994partial,
      turchin1998quantitative, wikle2003hierarchical,
      hooten2008hierarchical, wikle2010general,
      hooten2013computationally, williams2017integrated}. During
    diffusion, individual organisms are usually influenced by habitat
    type. Individuals move slowly through areas that contain necessary
    resources, and move quickly through areas that do
    not. \emph{Ecological diffusion} is a flexible diffusion model
    that accommodates this variation in motility by predicting animals
    will eventually accumulate in desirable habitats, and leave or
    avoid undesirable ones \citep{turchin1998quantitative,
      garlick2011homogenization, hefley2017when,
      williams2017integrated}. Specifically, ecological diffusion
    describes the population-level distribution that results from
    individual random walks, with individual movement probabilities
    determined by information on local habitat conditions
    \citep{garlick2011homogenization, hefley2017when,
      williams2017integrated}. Assuming no advection, ecological
    diffusion can be represented by the PDE
    \begin{linenomath}
      \begin{align}
        \frac{\partial u(\textbf{s},t)}{\partial t} =&
                                                       \bigg(\frac{\partial^2}{\partial
                                                       s_{1}^2}+\frac{\partial^2}{\partial
                                                       s_{2}^2}\bigg)[\mu(\textbf{s},t)u(\textbf{s},t)],
                                                       \label{eq:pde}
      \end{align}
    \end{linenomath}
    where $\frac{\partial u(\textbf{s},t)}{\partial t}$ represents the
    instantaneous change in abundance intensity over a continuous
    spatial domain with coordinates (e.g., latitude and longitude)
    $\textbf{s} \equiv (s_1,s_2)' \in \mathcal{S}$ during time $t$,
    $\bigg(\frac{\partial^2}{\partial
      s_{1}^2}+\frac{\partial^2}{\partial s_{2}^2}\bigg)$ is the
    differential (Laplace) operator, and $\mu(\textbf{s},t)$
    represents the diffusion coefficient that could vary in space and
    time. Ecological diffusion differs from other common
    reaction-diffusion models, in that it allows individual movement
    to be based on local conditions, rather than non-local conditions
    \citep[c.f., Fickian and plain
    diffusion;][]{garlick2011homogenization}. The mathematical driver
    for this difference is that the diffusion coefficient occurs on
    the inside of the two spatial derivatives rather than between them
    (e.g., Fickian:
    $\frac{\partial u}{\partial t} = \frac{\partial}{\partial x}
    \mu\frac{\partial}{\partial x}(u)$) or on the outside (e.g.,
    plain:
    $\frac{\partial u}{\partial t}=\mu \frac{\partial^2}{\partial
      x^2}(u)$), resulting in a much less smooth process, and
    motility-driven congregation to differ sharply between neighboring
    habitat types
    \citep{hooten2013computationally}. \cite{hefley2017when} recently
    described the advantages of ecological diffusion for modeling a
    spreading population including: its ability to connect
    spatio-temporal processes while providing a mechanism that
    captures transient dynamics, preventing animals from
    instantaneously accessing all high quality habitats; its relative
    simplicity compared to other mechanistic models; and its
    flexibility in being able to capture a wide range of
    spatio-temporal dynamics. For example, eq. \ref{eq:pde} can be
    further generalized to include growth models,
    \begin{linenomath}
      \begin{align}
        \frac{\partial u(\textbf{s},t)}{\partial t} =&
                                                       \bigg(\frac{\partial^2}{\partial
                                                       s_{1}^2}+\frac{\partial^2}{\partial
                                                       s_{2}^2}\bigg)[\mu(\textbf{s},t)u(\textbf{s},t)]+
                                                       f(u(\textbf{s},t),\textbf{s},t),
                                                       \label{eq:pde2}
      \end{align}
    \end{linenomath}
    incorporating Malthusian growth ($f(u(\textbf{s},t),\textbf{s},t)$
    $= \gamma(\textbf{s},t)u(\textbf{s},t)$), or logistic growth
    ($f(u(\textbf{s},t),\textbf{s},t)$
    $=\gamma(\textbf{s},t) (1-u(\textbf{s},t)/\kappa(\textbf{s},t)))$
    where $\gamma(\textbf{s},t)$ represents the instantaneous growth
    rate, and $\kappa(\textbf{s},t)$ represents equilibrium population
    size. In principle, each of the modeling components, including
    motility ($\mu(\textbf{s},t)$), growth ($\gamma(\textbf{s},t)$),
    and equilibrium density ($\kappa(\textbf{s},t)$) can depend on
    covariates that vary over space and time, although standard
    model-fitting considerations apply (i.e., parsimony) when
    tailoring these models to each system. We consider models that
    incorporate spatial covariates for diffusion,
    $g(\mu(\textbf{s}_i,t))=\textbf{X}\boldsymbol{\beta}$, and growth,
    $h(\gamma(\textbf{s}_i))=\textbf{W}\boldsymbol{\alpha}$, where $g$
    and $h$ are link functions, $\boldsymbol{\beta}$ and
    $\boldsymbol{\alpha}$ are parameters to be estimated, and
    \textbf{X} and \textbf{W} are matrices containing spatially
    referenced covariate values.

    Implementation of eqs. \ref{eq:pde} and \ref{eq:pde2} require
    numerical methods to solve the PDE. Finite differencing is a
    common method for solving PDEs, and is often used when PDEs are
    implemented within a Bayesian hierarchical framework
    \citep{wikle2010general}. Solving a PDE using finite differencing
    involves partitioning the spatial domain $\mathcal{S}$ into a grid
    \textbf{S} (\textbf{S}$\subseteq\mathcal{S}$) with $q$ cells and
    the temporal domain $\mathcal{T}$ into $r$ bins \textbf{T} of
    width $\Delta t$ (\textbf{T}$\subseteq\mathcal{T}$).  Simple
    finite-difference discretization results in the vector difference
    equation
    \begin{linenomath}
      \begin{align}
        \textbf{u}_t =& \textbf{H}(\boldsymbol{\alpha,\beta})
                        \textbf{u}_{t-1}+\textbf{H}(
                        \boldsymbol{\alpha,\beta})^{(\text{b})}
                        \textbf{u}_{t-1}^{(\text{b})},~~~~~~~~~~t=2,...,T
                        \label{eq:finitediff}
      \end{align}
    \end{linenomath}
    where \textbf{u}$_t \approx u(\textbf{s},t)$,
    \textbf{H}$(\boldsymbol{\alpha,\beta})$ is a sparse $q \times q$
    matrix with five non-zero diagonals accommodating diffusion
    parameters ($\boldsymbol{\beta}$) and growth parameters
    ($\boldsymbol{\alpha}$), and the superscript (b) represents
    conditions at the boundaries. To simplify notation in what
    follows, we assume \textbf{H} depends on diffusion and growth
    parameters, but drop the notation for $\boldsymbol{\alpha}$,
    $\boldsymbol{\beta}$. We also drop the notation for boundary conditions. The
    accuracy of the numerical approximation of $u(\textbf{s},t)$
    increases as the number of cells on the spatial grid increases and
    $\Delta t$ becomes small. For additional details on discretization
    of PDEs and applications of spreading populations, see
    \cite{wikle2006hierarchical}, \cite{hooten2008hierarchical}, and
    \cite{williams2017integrated}; both \cite{wikle2006hierarchical}
    and \cite{williams2017integrated} provide R code for
    implementation \citep[see][for ecological
    diffusion]{williams2017integrated}.

    \subsection*{Models of ecological diffusion and statistical uncertainty}
    Bayesian hierarchical models can be described in terms of three
    levels \citep{berliner1996hierarchical}. At the top level, a data
    model links the observed data and associated variation to latent
    ecological processes. Next, a process model describes the
    underlying ecological processes (i.e., spatio-temporal
    colonization dynamics). Finally, parameter models represent prior
    knowledge about the parameter inputs in the ecological process
    model and data model. This framework allows us to incorporate
    mathematical models that characterize spreading populations, such
    as the PDEs in eqs. \ref{eq:pde} or \ref{eq:pde2}, as process
    models within a statistical framework, permitting appropriate
    estimation of uncertainty at multiple levels
    \citep{wikle2003hierarchical, hooten2008hierarchical,
      wikle2010general, cressie2011statistics,
      hooten2013computationally, hefley2017when,
      williams2017integrated}. Using the discretized form of
    ecological diffusion in eq. \ref{eq:finitediff}, this framework is
    written hierarchically as
    \begin{linenomath}
      \begin{align}
        \begin{split}
          \textbf{Data Model:}~~~~~~~~~~~~~ y_t(\textbf{s}_i) &\sim
          [y_t(\textbf{s}_i)|n_t(\textbf{s}_i),\phi],~~~~~~~~~~~~~~~~~~~~~~~~~~~~~t=1,\ldots,T,
          \\
          \textbf{Process Models:}~~~~~~~~~~~~~~~~~~~   \textbf{n}_t  &\sim  [\textbf{n}_t|\textbf{u}_t,\nu], \\
          \textbf{u}_t &=
          \textbf{H}\textbf{u}_{t-1},~~~~~~~~~~~~~~~~~~~~~~~~~~~~~~~~~~~~~~~~~~~~t=2,\ldots,T,
          \\
          \textbf{u}_1 &= f(\boldsymbol{\zeta}) \\
          \textbf{Parameter Models:}~~~~~~~~~~~~~~~~~~~~
          \boldsymbol{\theta} & \sim
          [\phi,\nu,\boldsymbol{\alpha},\boldsymbol{\beta},\boldsymbol{\zeta}],
          \label{eq:bhm}
        \end{split}
      \end{align}
    \end{linenomath}
    where $y_t(\textbf{s}_i)$ represents data collected during
    discrete time $t$ at spatial location $\textbf{s}_i$, $[a|b]$
    represents the probability density (or mass) function of variable
    $a$ given variable $b$ \citep{gelfand1990sampling}, and
    $\textbf{n}_t \equiv
    (n_t(\textbf{s}_1),\ldots,n_t(\textbf{s}_n))'$. The initial
    condition for $\mathbf{u}_1$ must also be specified, and is
    represented as a function of (potentially vector valued)
    parameters $\boldsymbol{\zeta}$. Bayesian hierarchical models that
    incorporate PDE processes are flexible and can be modified to
    address the specifics of the study \citep{hefley2017when}. For
    example, a common specification of eq. \ref{eq:bhm} for discrete
    data (e.g., count data), consists of a binomial data model (i.e.,
    $y_t(\textbf{s}_i)\sim \text{Binomial}(n_t(\textbf{s}_i),\phi)$,
    where $n_t(\textbf{s}_i)$ is the true latent abundance, and $\phi$
    is the detection probability), and a Poisson process model (i.e.,
    $\textbf{n}_t \sim \text{Poisson}(\textbf{u}_t)$, in which case
    $\nu$ is not necessary). Other process models include
    negative-binomial or Conway-Maxwell Poisson distributions
    \citep[in which case, $\nu$ is a parameter that controls either
    overdispersion or underdispersion,
    respectively;][]{wu2013hierarchical}. Equation \ref{eq:bhm} can be
    further generalized to address error in discretization, or model
    uncertainty. For example,
    $\textbf{u}_t = \textbf{H}\textbf{u}_{t-1} +
    \boldsymbol{\epsilon}_t$, where
    $\boldsymbol{\epsilon}_t \sim \text{Normal}(\boldsymbol{0},
    \sigma^2\textbf{I})$, and \textbf{I} is the identity matrix
    \citep{wikle2010general}.

    Although discretization of the PDE (i.e., eq. \ref{eq:finitediff})
    provides a convenient form that results in a series of matrix
    equations, it is important to note that the theoretical
    foundations for this model are based in continuous time and space,
    and discretization provides only an approximate solution that may
    contain error. Coarser discretizations are more likely to contain
    larger error. Further, maintaining the connection to the PDE
    defined in continuous time and space (as we do in our specific
    application to sea otters, below; eq. \ref{eq:seaottermodel}) is
    advantageous for development and facilitation of numerical
    techniques for efficient implementation \citep[e.g.,
    homogenization;][]{garlick2011homogenization,
      hooten2013computationally, hefley2017when}.

    \subsection*{Forecast distribution}
    Forecasting the ecological process and associated uncertainty is
    necessary for optimal dynamic survey design. That is, we seek the
    probability distribution of the true state at the future point in
    time when data will be collected, conditional on the data we
    collected in the past \citep[i.e., the forecast distribution, or
    the predictive process distribution,
    \emph{sensu}][]{hobbs2015bayesian}. The forecast distribution is
    defined as
    \begin{linenomath}
      \begin{align}
        \begin{split}
          [\textbf{u}_{T+1}|\textbf{y}_1,\ldots,\textbf{y}_T] &
          =\int\ldots\int[\textbf{u}_{T+1}| \textbf{u}_{T},
          \boldsymbol{\theta}][\textbf{u}_{1},\ldots,\textbf{u}_{T},
          \boldsymbol{\theta}|\textbf{y}_{1},
          \ldots,\textbf{y}_{T}]d\boldsymbol{\theta}d\textbf{u}_{1}
          \ldots d\textbf{u}_{T}.
        \end{split}
                 \label{eq:predprocdist}
      \end{align}
    \end{linenomath}
    The Bayesian hierarchical model described in eq. \ref{eq:bhm}
    provides straightforward calculation of the forecast distribution.
    Obtaining $[\textbf{u}_{T+1}|\textbf{y}_1,\ldots,\textbf{y}_T]$ is
    as simple as changing the range of the index for $t$ in
    eq. \ref{eq:bhm} to $t=2,...,T+1$, and sampling
    $\textbf{u}^{(k)}_{T+1}$ on each $k=1,\ldots,K$ iteration of an
    MCMC algorithm \citep{tanner1996tools, hobbs2015bayesian}. The
    posterior predictive distribution can then be easily obtained from
    the forecast distribution using two additional steps; first sample
    $\textbf{n}^{(k)}_{T+1} \sim
    [\textbf{n}_{T+1}|\textbf{u}^{(k)}_{T+1},\nu^{(k)}]$. Then sample
    $\textbf{y}^{(k)}_{T+1} \sim
    [\textbf{y}_{T+1}|\textbf{n}^{(k)}_{T+1},\phi^{(k)}]$ for all $k$
    in $K$ to obtain
    $[\textbf{y}_{T+1}|\textbf{y}_1,\ldots,\textbf{y}_T]$. The
    forecast distribution and posterior predictive distribution can
    then be used to select a survey design that is optimal with
    respect to a design criterion.

    \subsection*{Design criteria}
    Design criteria are mathematical representations of the objectives
    investigators seek to achieve by collecting data
    \citep{williams2016combining}. As such, design criteria are
    specific to each study. However, a common objective of collecting
    data for many studies is to reduce the uncertainty associated with
    ecological forecasts/predictions. That is, choose a survey design
    $d$ that allows us to minimize the uncertainty associated with
    $[\textbf{u}_{T+1}|\textbf{y}_1,\ldots,\textbf{y}_T]$, or some
    derived parameter of $\textbf{u}_{T+1}$ (e.g.,
    $\sum_{i=1}^nu_{i,T+1}$, the expected total abundance during time
    $T+1$).  Several authors have discussed specific design criteria
    \citep[e.g., ][]{wikle1999space, berliner1999statistical,
      wikle2005dynamic, le2006statistical, hooten2009optimal}, as well
    as efficient methods for estimating them (e.g., Kalman
    filters). Here, we consider choosing a design that minimizes the
    uncertainty of $u_{\text{total},T+1}=\sum_{i=1}^n u_{i,T+1}$, the
    sum of the dynamic spatio-temporal process representing abundance
    intensity in future years. Specifically, the design criterion we
    consider is
    \begin{linenomath}
      \begin{align}
        q_d
        &=\frac{1}{K}\sum_{k=1}^K\bigg(u_{total,T+1,d}^{(k)}-\frac{1}{K}\sum_{k=1}^Ku_{total,T+1,d}^{(k)}\bigg)^2,
          \label{eq:q}
      \end{align}
    \end{linenomath}
    where $k=1,\ldots,K$ corresponds to the $k^{\text{th}}$ MCMC iteration, and
    $u_{total,T+1,d}^{(k)}$ is the sum of the forecasted process in
    time $T+1$, estimated using real data,
    $\textbf{y}_1,\ldots,\textbf{y}_T$, \emph{and} future data,
    $y_{T+1,d}$. Obviously, future data are unavailable prior to the
    survey. Lacking such data, one approach is to use the mean of the
    posterior predictive distribution as a surrogate for future data,
    and assume it represents the true data that remain to be
    collected. This technique, known as \emph{imputation}, may not
    accommodate the proper uncertainty associated with data
    collection. Another technique, known as \emph{multiple
      imputation}, helps to account for the uncertainty associated
    with the modeled data that we intend to use for identifying
    optimal survey designs \citep[][Scharf et al. 2017 \emph{In
      Review}]{rubin1996multiple, hooten2017animal}.

    \subsubsection*{Multiple imputation}
    Implementing multiple imputation within a Bayesian model using
    MCMC is straightforward \citep{hooten2017animal}. First, the model
    is fit using the original data,
    $\textbf{y}_1,\ldots,\textbf{y}_T$. Second, $K$ posterior
    predictive realizations of future data $\textbf{y}^{(k)}_{T+1}$
    are sampled for MCMC samples $k=1,\ldots,K$, using the methods
    described in \emph{Forecast distribution}, above. Third, the model
    is re-fit using a modified MCMC algorithm. Instead of conditioning
    only on the fixed data, $\textbf{y}_1,\ldots,\textbf{y}_T$, on the
    $k^{\text{th}}$ iteration of the MCMC algorithm, we use the fixed
    data \emph{and} $\textbf{y}^{(k)}_{T+1}$. Finally, we obtain
    posterior summaries for model parameters, and derived parameters
    including $u_{\text{total},T+1}$. The modified MCMC algorithm will
    integrate over the uncertainty in the true future data, and
    incorporate the uncertainty in the inference for the model
    parameters \citep{hooten2017animal}.

    Given the Bayesian hierarchical model described in
    eq. \ref{eq:bhm}, the forecast distribution described in
    eq. \ref{eq:predprocdist} (and the associated posterior predictive
    distribution), and a design criterion described in eq. \ref{eq:q},
    pseudo-code for combining animal movement models and survey design
    to identify the optimal monitoring of a spreading population is
    provided in Box 1.

    \fbox{\begin{minipage}[t]{.9\textwidth} \textbf{Box
          1}. Pseudo-code for combining animal movement models and
        survey design to identify the optimal monitoring of a
        spreading population.
        \begin{enumerate}
        \item Fit a model (i.e., eq. \ref{eq:bhm}) with baseline data
          $\textbf{y}_{1},\ldots,\textbf{y}_{T}$.

        \item Forecast $\textbf{u}^{(k)}_{T+1}$ for all $k=1,\ldots,K$
          MCMC samples using eq. \ref{eq:predprocdist}.

        \item Sample $K$ posterior predictive realizations of future
          data $\textbf{y}_{T+1}^{(k)}$ for $k=1,\ldots,K$ MCMC
          samples.

        \item Select a design $d$ that contains a subset of all
          possible survey locations in study area $\mathcal{D}$.

        \item Use multiple imputation to re-fit the model with
          baseline data $\textbf{y}_{1},\ldots,\textbf{y}_{T},$ and
          imputed data $\textbf{y}_{T+1,d}^{(k)}$, where
          $\textbf{y}_{T+1,d}^{(k)}$ are imputed for locations defined
          by design $d$.

        \item Calculate
          $u^{(k)}_{\text{total},T+1,d}=\sum_{i=1}^n
          u^{(k)}_{i,T+1,d}$ from the model fit in step 5.

        \item Use $u^{(k)}_{\text{total},T+1,d}$ to calculate
          eq. \ref{eq:q} from the text.

        \item Repeat steps 1-7 for all designs under consideration,
          and identify the design that minimizes $q_d$.
        \end{enumerate}
      \end{minipage}}

    \vspace{5mm}
    \noindent After the optimal design has been identified, the new
    data, $\textbf{y}_{T+1,d}$, can be collected, the model can be
    subsequently re-fit using the new data, ecological learning can be
    assessed by comparing the previous model fit to the new model fit,
    and the procedure can be repeated to identify the optimal design
    for time $T+2$. In the next section, we apply this general
    procedure to identify optimal transects to survey for estimating
    the distribution, abundance, and colonization dynamics of sea
    otters in Glacier Bay.

    \section*{Application: sea otters in Glacier Bay}

    We used the general framework described above to identify an
    optimal dynamic survey design for sea otters in Glacier Bay. We
    used baseline data to develop a Bayesian hierarchical model of
    population spread, with a process model tailored from the general
    ecological diffusion PDE described in eq. \ref{eq:pde2}. We then
    use our model to forecast abundance and distribution to a future
    monitoring period. Finally, we select a design that is optimal
    with respect to the forecast distribution, and a design criterion
    motivated by minimizing process prediction uncertainty.

    \subsection*{Baseline data}
    Sea otter occupancy and abundance data have been collected over a
    20-year period between 1993 and 2012. A detailed description of
    the methods that were used for collecting data are provided in
    \cite{bodkin1999aerial} and
    \cite{williams2017integrated}. Briefly, a \emph{design-based
      survey} was conducted eight times (1999--2004, 2006, 2012), and
    a \emph{distributional survey} was conducted eight times (1993,
    1995--1998, 2005, 2009, 2010). The design-based survey consisted
    of observers flying in aircraft piloted along transects. The
    transects were systematically placed across Glacier Bay, with a
    random starting point. Observers flew along transects and recorded
    the number of sea otters observed within 400 m of the transect,
    and mapped the location of sea otters during observations. The
    distributional surveys consisted of observers flying in aircrafts
    that were piloted in close proximity to shorelines and islands,
    the preferred habitat of sea otters \citep[][Williams et
    al. \emph{In Review}]{williams2017integrated}. Pilots did not
    follow pre-determined routes during distributional surveys.  An
    additional data set was collected during the design-based survey
    to facilitate estimating detection probability
    \citep{williams2017integrated}.

    \subsection*{Statistical diffusion model and forecast}

    We tailored eq. \ref{eq:bhm} to the sea otter data following
    \cite{williams2017integrated} and Williams et al. (\emph{In
      Review}). Retaining connection to the continuous time,
    continuous space process model, we assumed
    \begin{linenomath}
      \begin{align}
        \begin{split}
          \textbf{Data Model:}~~~~~~~~~~~~~~~ y_t(\textbf{s}_i)
          &\sim \text{Binomial}(n_t(\textbf{s}_{i}),\phi), \\
          \textbf{Process Model:}~~~~~~~~~~~~~~n_t(\textbf{s}_i) &\sim
          \text{Poisson}(u_t(\textbf{s}_i)),
          \\
          \frac{\partial u(\textbf{s}_i,t)}{\partial t}
          &=\bigg(\frac{\partial^2}{\partial s_{1}^2}+
          \frac{\partial^2}{\partial s_{2}^2}\bigg)
          [\mu(\textbf{s},t)u(\textbf{s},t)]+\gamma(\textbf{s}_i)u(\textbf{s}_i,t),~~~~~~~~~t>1
          \\
          u(\textbf{s}_i,1) &= \frac{\tau
            e^{\frac{-|\textbf{s}_i-\textbf{d}|^2}{\kappa^2}}}
          {\int_Se^{\frac{-|\textbf{s}_i-\textbf{d}|^2}{\kappa^2}}ds},~~~~~~~~~~~~~~~~
          ~~~~~~~~~~~~~~~~~~~~~~~~~~~~~~~~~~~~~~~~~t=1  \\
          \text{log}(\mu(\textbf{s}_i))&=\beta_0 +
          \beta_1(\text{depth}(\textbf{s}_i)) +
          \beta_2(\text{dist}(\textbf{s}_i)) +
          \beta_3(\text{depth}(\textbf{s}_i) \times
          \text{slope}(\textbf{s}_i))
          \\
          &+\beta_4(\text{complexity}(\textbf{s}_i))
          \\
          \gamma(\textbf{s}_i)&=\alpha_0  \\
          \textbf{Parameter Models:}~~~~~~~~~~~~~~~~~~~~~~\phi&\sim
          \text{Beta}(1,1)
          \\
          \boldsymbol{\beta}&\sim \text{Normal}(\boldsymbol{0},1.5^2\textbf{I})  \\
          \alpha & \sim \text{Normal}(0,1.5^2)  \\
          \kappa & \sim \text{Normal}^+(5,0.001) \\
          \tau & \sim \text{Normal}^+(500,10)
        \end{split}
                 \label{eq:seaottermodel}
      \end{align}
    \end{linenomath}
    where $y_t(\textbf{s}_i)$ were sea otter count data within a
    400$\times$400 m grid cell centered at location $\textbf{s}_i$
    during time $t$, $n_t(\textbf{s}_i)$ was the true latent
    abundance, $\phi$ was the individual sea otter detection
    probability, and $u_t(\textbf{s}_i)$ was the dynamic
    spatio-temporal process (abundance intensity) when data were
    collected during time $t$. We used a scaled Gaussian kernel for
    our initial condition for abundance intensity, with two parameters
    $\boldsymbol{\zeta} \equiv (\tau,\kappa)'$, controlling the height
    and spread of the kernel, respectively, around an epicenter
    $\textbf{d}$. We used a log-linear relationship between motility
    and four spatial habitat covariates that we hypothesized affect
    sea otter motility. The covariates were ocean depth (an indicator
    of depth$<$40 m), distance to shore, slope of the ocean
    floor, and an index for shoreline complexity that was calculated
    by summing the number of shoreline grid cells that were within
    1,000 m of each grid cell. We used the interaction between depth
    and slope because the slope of the ocean floor may
    only be important if it is shallow enough for sea otters to reach
    it during feeding dives. We assumed the growth rate was constant
    across space and time. We used vague prior distributions for all
    parameters except for the initial condition parameters, $\tau$ and
    $\kappa$, which we parameterized based on observations of sea
    otters during the first year of monitoring, where
    $\text{Normal}^+$ represents the zero-truncated normal
    distribution.

    We fit the model described in eq. \ref{eq:seaottermodel} to the
    baseline data using a custom MCMC algorithm written in R version
    3.3.2 \citep{r332} and C++. For each model fit, we obtained two
    chains of 20,000 MCMC draws and discarded the first 5,000. We
    examined convergence using trace plots and Gelman-Rubin
    diagnostics. To facilitate computation, we used homogenization to
    implement the model \citep{garlick2011homogenization,
      hooten2013computationally, hefley2017when,
      williams2017integrated}. We used regularization combined with
    k-fold cross-validation to conduct model selection, and assessed
    goodness of fit using Bayesian p-values (see Williams et
    al. \emph{In Review} for details). We then estimated the forecast
    distribution for $T+5=2017$, because the last time sea otter data
    were collected was $T=2012$ (Fig. \ref{fig:forecast}).

    \subsection*{Optimal design}
    \subsubsection*{Potential survey transects}
    To identify the set of all potential transects that could be
    surveyed, we partitioned Glacier Bay into a regular grid of
    400$\times$400 m cells. We selected 400 m as the unit of length
    for two reasons. First, this partitioning assisted with
    computation, because computation at a finer resolution became
    prohibitive. Second, 400$\times$400 m represented the scale at
    which the baseline data were collected. After partitioning Glacier
    Bay into 400$\times$400 m grid cells, there were 170 potential
    transects (running West to East) from which we could select a
    sampling design. This resulted in $\binom{170}{n}$ unique possible
    designs that could be considered, where $n$ is the number of
    transects that could be flown during a survey.

    \subsubsection*{Selecting an optimal design}
    We considered a sample size of $n=20$ transects to be observed for
    our monitoring design, which is approximately the maximum number that can be
    flown in one day. This resulted in a total number of possible
    designs that was much larger than one trillion. It is not feasible
    to calculate the design criterion $q_d$ for all possible unique
    designs, thus we considered an approach based on improving
    efficiency relative to a random selection of transects. First, we
    selected a large number of different designs, $d$, uniformly at
    random and calculated the design criterion $q_d$ for each design
    using eq. \ref{eq:q}. Fitting the sea otter model described in
    eq. \ref{eq:seaottermodel} to the baseline data described above,
    and calculating $q_d$ for one design required approximately 5.4
    hours to obtain 20,000 MCMC samples. To facilitate fitting a large
    number of different random designs, we used the Amazon Elastic
    Compute Cloud (Amazon EC2\textregistered, instance: Linux
    m4.16xlarge; with 64 vCPUs) to calculate $q_d$ for 64 different
    random designs in parallel. We then compared $q_d$ among all 64
    designs, and selected the design that minimized $q_d$. A histogram
    of the $q_d$ values for all 64 random designs we examined is shown
    in Fig. \ref{fig:q}.

    After we identified the optimal set of random transects, we
    further improved the design using an exchange algorithm
    \citep{royle1998algorithm}. That is, we sequentially exchanged
    each of the 20 transects with their neighbors (one transect above
    it, and one transect below it), and recalculated $q_d$ after the
    exchange. This required re-fitting the model with the inclusion of
    a neighboring transect and the exclusion of the original
    transect. If the exchange improved $q_d$, we retained the new
    transect in place of the old transect. Then, the next transect on
    the list was exchanged. The process repeated until the design
    criterion could not be improved through exchange. Because each
    exchange requires re-fitting the model, and it must occur
    sequentially (except for examining the two immediate neighbors,
    which can occur in parallel), this required a sequence of several
    model fits. However, in practice, convergence to the optimal
    survey design occurs with relatively few exchanges using this
    approach. The sea otter survey required six exchanges before $q_d$
    could no longer be improved through exchange.

    \subsection*{Results}
    The posterior mean abundance estimates of sea otters in 2017 were
    similar among all designs (mean = 9,430; range = 9,250--9,770),
    suggesting mean abundance estimates were not sensitive to the
    choice of designs we considered. However $q_d$ values ranged from
    66,685 (best) to 88,948 (worst) and averaged 76,680
    (Fig. \ref{fig:q}). Thus, the $q_d$ value of the optimal design
    improved by 13\% when compared to the average $q_d$ value of all
    other designs we considered.  The optimal survey design is shown
    in Fig. \ref{fig:OptimalDesignPlot}.

    \section*{Discussion}
    How to best use available resources to monitor ecological
    processes for conservation, management, and ecological insight
    remains a critical area of scientific investigation
    \citep{nichols2006monitoring}. Probabalistic (i.e., design-based)
    surveys have been used widely in ecology, and can provide data
    that result in objective, unbiased estimates of abundance
    \citep{cochran2007sampling, thompson2012sampling}. However, when
    financial resources limit the effort that can be devoted to
    collecting data, classical design-based inference may result in
    estimates that are insufficiently precise for management or
    conservation (e.g., sea otters in Glacier Bay). The situation
    becomes more accute for populations that are spreading in space
    through time. Alternatively, optimal dynamic survey designs allow
    managers and scientists the ability to extract the most
    information out of the data they can afford to collect. Further,
    dynamic survey designs better allow for the observation of
    dynamically evolving spatio-temporal processes, and ultimately
    result in higher quality data \citep{wikle1999space,
      wikle2005dynamic, hooten2009optimal}.

    Optimal dynamic survey designs are becoming widespread in
    atmospheric and environmental studies. However, they have been
    used in relatively few long-term ecological studies
    \citep{hooten2009optimal}. While model-based inference has become
    ubiquitous in ecology, survey design and modeling are usually
    developed independently of each other. By explicitly linking
    survey design, and the models that will be fit to future data, we
    gain the ability to employ more sophisticated ecological models
    that ultimately contain less uncertainty
    \citep{hooten2009optimal}.

    We described a general, cohesive framework for modeling and
    monitoring population-level animal movement that explicitly links
    survey design, data collection, and monitoring objectives. The
    generality of this framwork stems from the flexibility of
    hierarchical statistical models to draw conclusions from data that
    arise from complex ecological processes, the flexibility of PDEs
    (specifically, ecological diffusion) to capture a wide range of
    spatio-temporal dynamics, and the ability to tailor design
    criteria to meet the objectives of each unique study. We applied
    the framework to identify an optimal dynamic survey design for sea
    otters in Glacier Bay.  Sea otters have been identified as a vital
    sign for Glacier Bay. Vital-sign monitoring is used to track
    specific ecosystem processes that are selected to represent the
    overall health or condition of park resources, known or
    hypothesized effects of stressors, or elements that have important
    human values. Inference that results from monitoring is then used
    by employees and partners to support management decision-making,
    park planning, research, education, and public understanding of
    park resources. Thus, a survey design that results in precise,
    rigorous, and honest estimates of abundance, distribution, and
    colonization dynamics is required. We examined a monitoring
    scenario in which available funding permitted surveying 20 of the
    170 potential transects that partition Glacier Bay. Generally,
    posterior mean estimates of sea otter expected abundance were
    similar among the designs we considered; all designs predicted
    approximately 9,500 sea otters in 2017. However, the uncertainty
    associated with these predictions varied widely among designs. The
    optimal design reduced prediction uncertainty by 13\% compared to
    the mean of all the random designs that were considered
    (Fig. \ref{fig:q}). The dynamic survey designs employed for sea
    otters surveys here, are applicable to any type of aerial survey
    method used for sea otters, including aerial observations where
    observers count sea otters from an aircraft
    \citep{bodkin1999aerial}, or aerial photographs
    \citep{williams2017estimating}.

    The design criterion we employed, chosen by the National Park
    Service, is a measure of the prediction uncertainty of the
    expected abundance of sea otters in Glacier Bay (i.e., how many
    sea otters are there next year).  Many choices of
    design criteria are possible, and depend on the objectives of the
    study. The explicit choice of a design criterion pairs survey
    design with the motives of a decision maker in a decision
    theoretic framework \citep{wald1950statistical,
      savage1954foundations, williams2016combining}. This pairing is
    natural in monitoring for ecology because data are often collected
    with the explicit purpose to inform both models and
    decisions. \citet[][p. 668]{nichols2006monitoring} state
    ``targeted monitoring is defined by its integration into
    conservation practice, with monitoring design and implementation
    based on \emph{a priori} hypotheses and associated models of
    system responses to management.'' Thus, the framework we present
    is directly amenable to \emph{targeted monitoring}, sensu
    \cite{nichols2006monitoring}, due to the explicit incorporation of
    a design criterion. Further, by selecting a design criteria
    focused on minimizing structural (i.e., multi model) uncertainty,
    or the uncertainty associated with management actions, the
    framework becomes amenable to adaptive resource management
    \citep[e.g.,][]{johnson1997uncertainty}, and our framework
    provides an efficient method for achieving targeted monitoring for
    conservation. That is, it is a method for explicitly focusing
    monitoring efforts on crucial information needs in the
    conservation process, and therefore, the effectiveness of
    conservation can be greatly increased
    \citep{nichols2006monitoring}.

    Extentions of dynamic survey designs include hybrid survey
    designs. Hybrid survey designs combine classical survey techniques
    (e.g., random sampling) with dynamic survey designs to identify an
    optimal dynamic survey design \citep{hooten2009optimal,
      hooten2012optimal}. Hybrid survey designs are advantageous
    because they leverage the benefits of traditional survey
    techniques (e.g., generally more convenient, economically
    feasible, and computationally inexpensive), with the benefits of
    dynamic survey designs \citep[e.g., optimal efficiency, capture
    spatio-temporal evolution in a process, flexibility to add or
    remove monitoring locations as budgets
    change;][]{hooten2009optimal}. When hybrid survey designs contain
    a design-based sampling component, the design-based data can stand
    alone to obtain design-based estimates of abundance, which
    provides desirable operating characteristics \citep[e.g., unbiased
    estimation;][although at a cost in precision]{cochran2007sampling,
      thompson2012sampling}.

    Finally, spreading populations are ideal candidates for dynamic
    survey designs because spreading populations have significant
    spatio-temporal interactions that are difficult to observe using
    traditional survey designs. The spatio-temporal processes that
    regulate population spread are often of ecological interest
    \citep[e.g., processes that influence species invasions,
    mesopredator release, (re)establishment of apex
  predators;][]{williams2017integrated}. When baseline data exist to
    develop appropriate models of population spread, implementing
    dynamic survey designs for future data collection provide an
    opportunity to maximize efficiency in learning about these
    spatio-temporal processes \citep{wikle1999space}. When resources
    are limited, as they always are, the efficient use of monitoring
    is vital to successful conservation \citep{nichols2006monitoring}.

    \section*{Acknowledgments}

    Funding was provided from the National Park Service Inventory and
    Monitoring Program, Glacier Bay National Park Marine Management
    Fund, and NSF DMS 1614392. Heather Coletti, Dan Esler, Dan Monson,
    and John Tipton provided technical and logistical support. Data
    were collected by James Bodkin, Janet Doherty, Dan Monson, and Ben
    Weitzman. Pat Kearney and Andy Harcombe piloted sea otter survey
    airplanes.  Any use of trade, firm, or product names is for
    descriptive purposes only and does not imply endorsement by the
    U.S. Government.

    \bibliographystyle{ecology} \bibliography{references}

    \clearpage

    \paragraph{Figure 1}
    Schematic of optimal dynamic survey design.

    \paragraph{Figure 2}
    Forecasted mean of dynamic spatio-temporal process
    ($u_{2017}(\textbf{s})$) representing abundance intensity of sea
    otters in Glacier Bay National Park, Alaska. Units are mean sea
    otters per 400 m$^2$

    \paragraph{Figure 3}
    Histogram of $q_d$ values from 64 randomly selected designs (gray)
    and the optimal design (black), each design containing 20 randomly
    selected transects to be flown over Glacier Bay National Park in
    the upcoming survey year. The design criterion $q_d$ was
    calculated using eq. \ref{eq:q} from the text, and corresponds to
    reducing uncertainty in the forecast distribution of mean total
    abundance of sea otters in the future year. The best random design
    had $q_d=57,439$ (dark gray), and was improved to $q_d=55,261$
    (black) using an exchange algorithm. The mean value of $q_d$ for
    the 64 random transects equaled 62,804 (vertical line).

    \paragraph{Figure 4}
    Optimal dynamic survey design for sea otters in Glacier Bay
    National Park, 2017.


    \begin{center}
      \begin{figure}[!ht]
        \includegraphics[width=\linewidth]{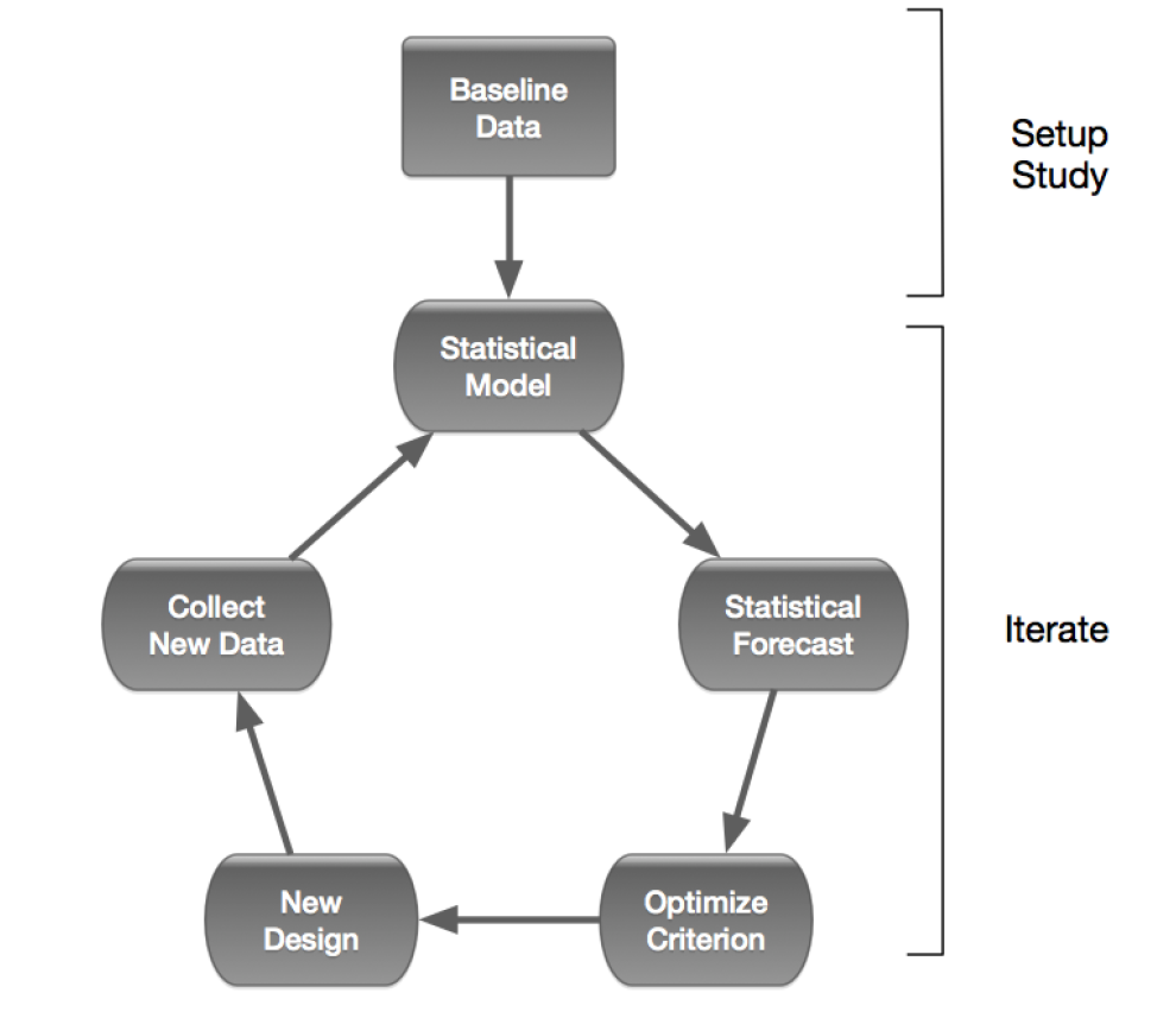}
        \caption{}
        \label{fig:schematic}
      \end{figure}
    \end{center}


    \begin{center}
      \begin{figure}[!ht]
        \includegraphics[width=\linewidth]{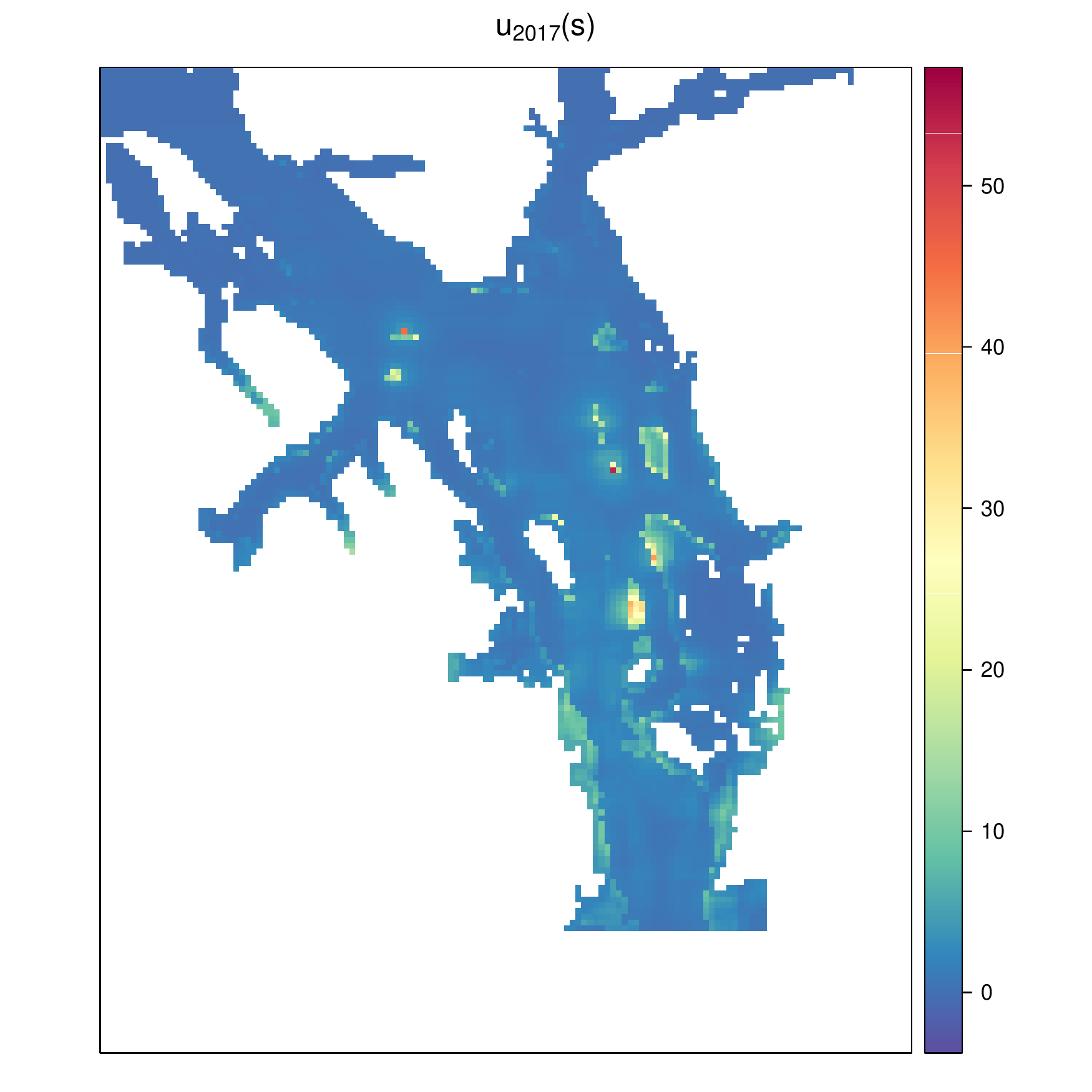}
        \caption{}
        \label{fig:forecast}
      \end{figure}
    \end{center}


    \begin{center}
      \begin{figure}[!ht]
        \includegraphics[width=\linewidth]{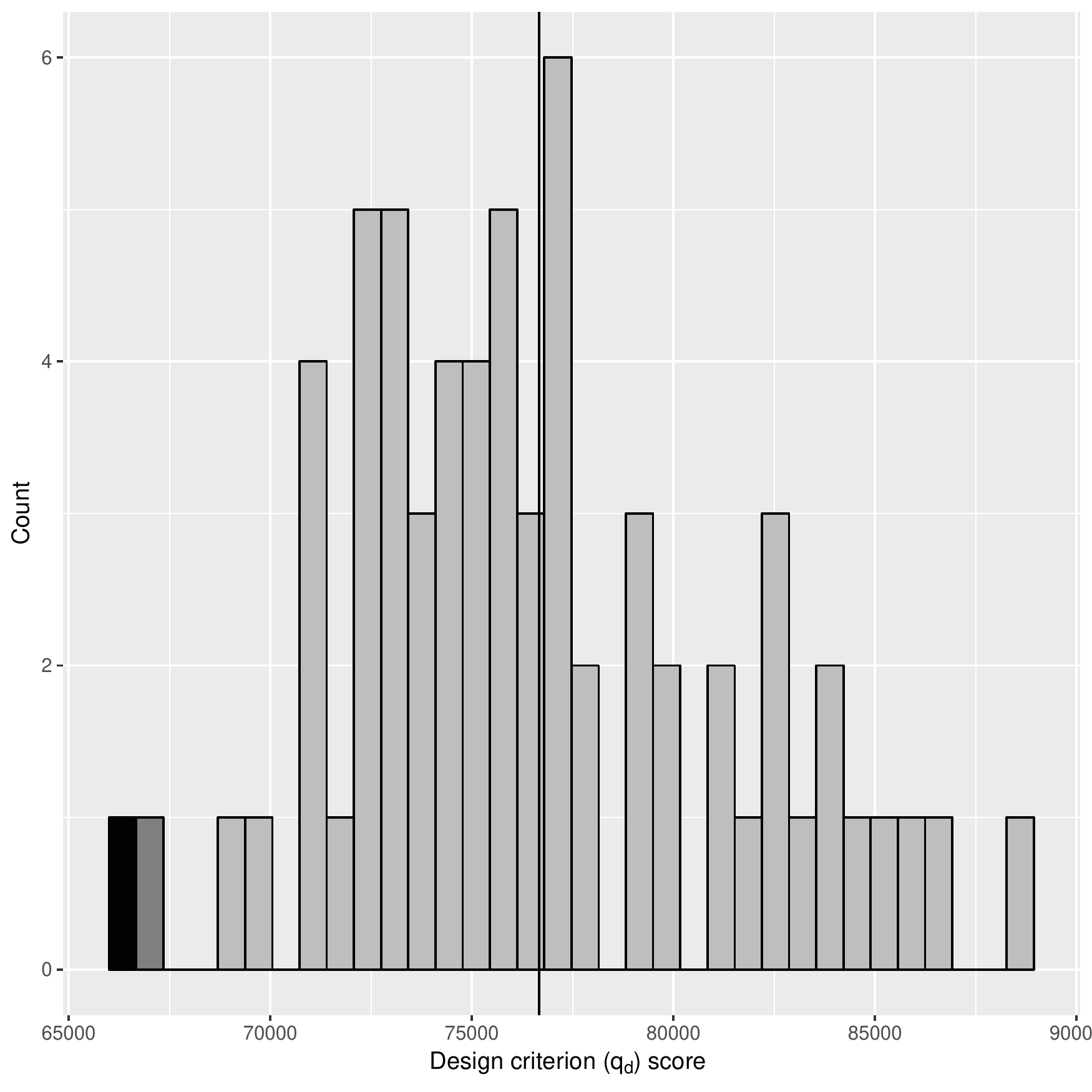}
        \caption{}
        \label{fig:q}
      \end{figure}
    \end{center}


    \begin{center}
      \begin{figure}[!ht]
        \includegraphics[width=\linewidth]{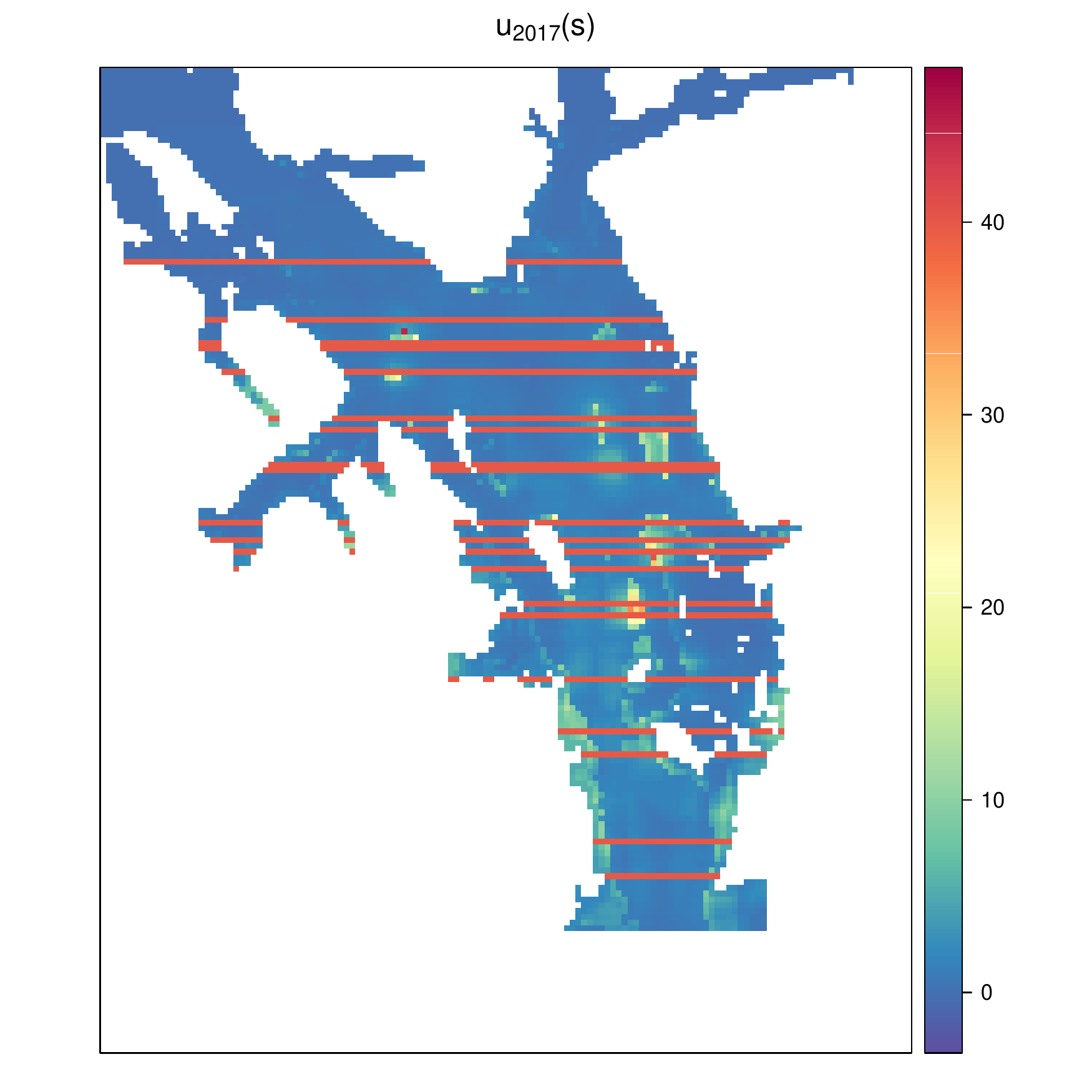}
        \caption{}
        \label{fig:OptimalDesignPlot}
      \end{figure}
    \end{center}

  \end{flushleft}
\end{spacing}
\end{document}